\documentstyle[prb,twocolumn,epsf,floats,aps]{revtex}
\begin{document}
\draft

\twocolumn[\hsize\textwidth\columnwidth\hsize\csname @twocolumnfalse\endcsname

\title{The sign problem in Monte Carlo simulations of frustrated
quantum spin systems}

\author{Patrik Henelius$^1$ and Anders W. Sandvik$^{2}$}

\address{$^1$ National High Magnetic Field Laboratory, 1800 East Paul
Dirac Dr., Tallahassee, Florida 32310 \\ $^2$ Department of Physics,
University of Illinois at Urbana-Champaign, 1110 West Green Street,
Urbana, Illinois 61801}

\date{\today}

\maketitle

\begin{abstract}
We discuss the sign problem arising in Monte Carlo simulations of
frustrated quantum spin systems. We show that for a class of
``semi-frustrated'' systems (Heisenberg models with ferromagnetic
couplings $J_z(r) < 0$ along the $z$-axis and antiferromagnetic
couplings $J_{xy}(r)=-J_z(r)$ in the $xy$-plane, for arbitrary
distances $r$) the sign problem present for algorithms operating in
the $z$-basis can be solved within a recent ``operator-loop''
formulation of the stochastic series expansion method (a cluster
algorithm for sampling the diagonal matrix elements of the power
series expansion of ${\rm exp}(-\beta H)$ to all orders). The solution
relies on identification of operator-loops which change the
configuration sign when updated (``merons'') and is similar to the
meron-cluster algorithm recently proposed by Chandrasekharan and Wiese
for solving the sign problem for a class of fermion models
(Phys. Rev. Lett. {\bf 83}, 3116 (1999)). Some important expectation values,
e.g., the internal energy, can be evaluated in the subspace with no
merons, where the weight function is positive definite.  Calculations
of other expectation values require sampling of configurations with
only a small number of merons (typically zero or two), with an
accompanying sign problem which is not serious. We also discuss
problems which arise in applying the meron concept to more general
quantum spin models with frustrated interactions.
\end{abstract}


\vskip2mm]

\section{INTRODUCTION}
Recently, there have been several significant developments of more
efficient Monte Carlo methods for interacting quantum many-body
systems.\cite{reviews} The Trotter decomposition formula
\cite{suzuki1,worldline} has traditionally been used as a starting
point for finite-temperature simulation algorithms, such as the
worldline \cite{worldline} and fermion determinant \cite{determinant}
methods. It introduces a systematic error that can be removed only by
carrying out simulations for several different imaginary time
discretizations $\Delta\tau$ and subsequently extrapolating to
$\Delta\tau=0$. Such extrapolations are not necessary with the
stochastic series expansion method, \cite{sse1,sse2,sse3} which is
based on sampling the power series expansion of exp$(-\beta H)$ to all
orders and is related to a less general method proposed much earlier
by Handscomb.\cite{handscomb,lyklema,lee} Results that are exact to
within statistical errors can also be directly obtained with recent
worldline \cite{prokofev,beard,irsse} and fermion determinant
\cite{rombouts} algorithms formulated in continuous imaginary
time. Even more significant than the elimination of the Trotter
decomposition error are generalizations to the quantum case
\cite{evertz,kawashima,beard,sse3} of cluster algorithms developed
for classical Monte Carlo.\cite{swendsen} These
``loop algorithms'' (so called because the clusters are loops on a
space-time lattice) can reduce the autocorrelation times by orders of
magnitude and enable highly accurate studies of systems in parameter
regimes where previous algorithms encountered difficulties due to long
autocorrelation and equilibration times.

In spite of these developments, the class of models which can be
studied using quantum Monte Carlo methods is still severely restricted 
due to the ``sign problem'',\cite{loh,miyashita} i.e., the
non-positive-definiteness of the weight function that can arise in
transforming a quantum problem into a form resembling a classical
statistical mechanics problem. There are two classes of systems for
which this issue is particularly pressing --- interacting fermions in
more than one dimension and quantum spin systems with frustrated
interactions (in any number of dimensions). For fermions in one
dimension, and hopping between nearest-neighbor sites only, the sign
problem can be avoided because the fermion anticommutation relations
do not come into play (other than introducing a hard-core constraint)
in the one-dimensional real-space path integral. In two or more
dimensions (or even in one dimension if hopping further than between
nearest neighbors is included), permutations of fermions during the
propagation in imaginary time lead to a mixed-sign path integral which
typically cannot be efficiently evaluated using Monte Carlo methods.
The sign problem can be avoided with the fermion determinant algorithm
in special cases, such as the half-filled Hubbard model (because of
particle-hole symmetry),\cite{determinant} but in other cases
simulations are restricted to high temperatures and/or small system
sizes.\cite{loh} For frustrated spin systems the source of the sign
problem is different. A minus sign appears for every event in the path
integral in which two antiferromagnetically interacting spins are
flipped.\cite{miyashita} This causes an over-all minus sign if the
total number of spin flips is odd, which can be the case, e.g., for a
triangular lattice or a square lattice with both nearest- and
next-nearest-neighbor interactions. Simulations of quantum spin
systems are therefore restricted to models with no frustration (in the
off-diagonal part of the Hamiltonian), such as ferromagnets, or
antiferromagnets on bipartite lattices.

A promising approach to solving the sign problem was recently
suggested by Chandrasekharan and Wiese. \cite{wiese} They considered a
system of spinless fermions on a two-dimensional square lattice within
the context of the worldline loop algorithm.\cite{evertz} They showed
that, for this particular model and for a certain range of
nearest-neighbor repulsion strengths, the properties of the loops can
be used to eliminate the sign problem. Flipping a loop can change the
number of fermion permutations from odd to even, or vice versa,
thereby also changing the over-all sign of the configuration. Such
sign-changing loops are called ``merons''.  The magnitude of the
weight is not affected by flipping a meron and therefore all
configurations with one or more merons cancel in the partition
function. The subspace of zero merons is positive definite and can be
sampled without a sign problem. Typical operator expectation values of
interest also contain contributions from configurations with two
merons which therefore also have to be included in the simulation and
introduce a ``mild'' sign problem. The relative weights of the zero-
and two-meron subspaces to be sampled can further be chosen in an
optimum way using a reweighting technique, which further reduces the
sign problem.

In this paper we explore an analogous method for solving the sign
problem for frustrated quantum spin models. We consider the
operator-loop formulation of the stochastic series expansion
method,\cite{sse3} in which sequences of two-spin operators are
sampled by forming clusters (loops) of operators that can be
simultaneously updated without changes in the weight function.  The
updated clusters contain operators acting on the same spins, but
diagonal operators may be changed to off-diagonal ones and vice
versa. For a model with frustrated interactions an operator-loop
update can lead to a sign change. In analogy with Chandrasekharan and
Wiese\cite{wiese} we will refer to such sign-changing loops as
``merons''. The sign problem can be solved if the operator-loops for a
given configuration can be uniquely defined and the weight function is
positive definite in the configuration subspace containing no merons.
Unfortunately, we find that these criteria are in general difficult to
satisfy. Operator-loop algorithms with uniquely determined loops are
typically non-ergodic for frustrated systems, and with supplemental
local updates for ergodicity there are mixed signs in the zero-meron
subspace.  In fact, in such cases merons typically do not even exist,
i.e., none of the operator-loops can change the sign when flipped. We
have found only one spin system for which the sign problem can be
eliminated using merons --- the Heisenberg model with ferromagnetic
couplings $J_z(r) < 0$ along the $z$-axis and frustrated
antiferromagnetic couplings $J_{xy}(r)=-J_z(r)$ in the plane
perpendicular to this axis, i.e., the Hamiltonian
\begin{equation}
H = -\sum\limits_{i,j}J_{ij}[S^z_iS^z_j - \hbox{$1\over 2$}
(S^+_iS^-_j+S^-_iS^+_j)],
\label{subversivemodel}
\end{equation}
where $J_{ij} > 0$ and the range of the couplings is arbitrary. We
have implemented a meron algorithm for this model on a square lattice
with nearest- and next-nearest-neighbor couplings $J(1)$ and
$J(\sqrt{2})$. Standard algorithms for this model have a severe sign
problem when using the $z$-axis as the quantization axis, however, it
can be avoided by a simple rotation to the $x$-representation.  Using
the SSE algorithm and the meron concept, the sign problem can be
eliminated also in the $z$ representation. With both representations
accessible in simulations, correlation functions both parallel and
perpendicular to the $z$ direction can be easily evaluated.

The model, Eq.~(\ref{subversivemodel}), can be mapped onto a hard-core
boson model with attractive interactions and frustrated
hopping. Frustration in the potential energy has been investigated in
this context as a possible mechanism to render a disordered bosonic
ground state.\cite{murthy} Frustration in the hopping [the $xy$-term
in Eq.~(\ref{subversivemodel})] should decrease the tendency to
forming off-diagonal long-range order and could then lead to a normal
fluid (non-superfluid). However, the highly symmetric case considered
here has a trivial, ordered ground state; the fully polarized
ferromagnetic state (corresponding to a completely filled lattice of
hard-core bosons; a trivial case of normal solid). Effects of
frustration only come into play at finite temperature, where the model
is different from the corresponding isotropic ferromagnet (on non
frustrated, bipartite lattices the two models are equivalent, since
the sign of the $xy$-term can be switched by a spin rotation on one of
the sublattices).

Although we have not been able to solve the sign problem for other
cases, such as the Heisenberg model with completely antiferromagnetic
interactions [$J_z(1) = J_{xy}(1) > 0$ and $J_z(\sqrt{2}) =
J_{xy}(\sqrt{2})>0$], our work nevertheless gives some further
insights into the meron concept and what is required in order to solve
the sign problem for arbitrary couplings.

The outline of the rest of the paper is the following: In Sec.~II we
review the basics of the stochastic series expansion method and
discuss operator-loop updating schemes for both ferromagnetic and
antiferromagnetic couplings. In Sec.~III we present the solution of
the sign problem for the $J_z(r)=-J_{xy}(r)$ model.  The reweighting
technique is analyzed in some detail in Sec.~IV. In Sec.~V we
discuss some simulation results for the semifrustrated model and make 
comparisons with the isotropic Heisenberg ferromagnet. We summarize our 
work in Sec.~VI.

\section{OPERATOR-LOOP ALGORITHM}

In this section we first briefly review the expansion underlying the
SSE method and then discuss the operator-loop updates used to
efficiently sample the expansion. We here assume a non-frustrated case
and postpone the discussion of the sign problem for frustrated models
to Sec.~III.  For definiteness we consider the $S=1/2$ Heisenberg
model
\begin{equation}
H=\pm J \sum_{\langle i,j \rangle} S_i\cdot S_j,
\end{equation}
where $\langle ij \rangle$ denotes a pair of nearest-neighbor spins on
a cubic lattice (in an arbitrary number of dimensions), and $J>0$.
Depending on the sign, the model is an antiferromagnet ($-$) or a
ferromagnet $(+)$. To construct the SSE configuration space the
Hamiltonian is rewritten as a sum of diagonal and off-diagonal
operators
\begin{equation}
H=-\frac{J}{2} \sum_{b=1}^M \left( H_{1,b}\mp H_{2,b} \right)+ C,
\end{equation}
where the index $b$ denotes an interacting spin pair (bond) ${\langle
i(b),j(b) \rangle}$ and $C$ is an irrelevant constant equal to $MJ/4$,
where $M$ is the total number of pairs of interacting spins.  The
bond-indexed operators are given by
\begin{eqnarray}
H_{1,b}&=& 2\left(\hbox{$\frac{1}{4}$}\mp
S_{i(b)}^zS_{j(b)}^z\right)\\ H_{2,b}&=& S_{i(b)}^+S_{j(b)}^- +
S_{i(b)}^-S_{j(b)}^+.
\end{eqnarray}
Note that the eigenvalues of both the diagonal ($H_{1,b}$) and the
off-diagonal ($H_{2,b}$) operators are $0$ and $1$, both for the
antiferromagnet and the ferromagnet. The partition function
$Z=\mbox{Tr} \{\exp(-\beta H)\}$ is expanded as \cite{sse1}
\begin{equation}
Z=\sum_{\alpha}\sum_{n=0}^{\infty} \frac{(-\beta)^n}{n!}
\langle\alpha|H^n|\alpha\rangle,
\label{part1}
\end{equation}
in the basis
$\{|\alpha\rangle\}=\{|S_1^z,S_2^z,\ldots,S_z^N\rangle\}$, where $N$
is the number of spins. Terms of order greater than 
$n\sim N\beta$
give an exponentially vanishing contribution and for the purpose of a
stochastic sampling the expansion can therefore be truncated at some
$n=L$ of this order without loss of accuracy (see, e.g.,
Ref.\onlinecite{sse2} for details on how to choose a sufficiently high
truncation power). Additional unit operators $H_{0,0} \equiv 1$ are 
introduced to rewrite Eq.~(\ref{part1}) as
\begin{equation}
Z=\sum_{\alpha}\sum_{S_L} {(\mp 1)^{n_2} (J\beta)^n (L-n)! \over 2^n
L! } \left\langle\alpha\left|\prod_{i=1}^L
H_{a_i,b_i}\right|\alpha\right\rangle,
\label{part2}
\end{equation}
where $S_L$ denotes a sequence of operator-indices:
\begin{equation}
S_L=(a_1,b_1)_1,(a_2,b_2)_2,\ldots,(a_L,b_L)_L,
\end{equation}
with $a_i\in \{1,2\}$ and $b_i\in \{1,\ldots,M\}$, or
$(a_i,b_i)=(0,0)$.  The number of non-(0,0) elements in $S_L$ is
denoted $n$, while $n_2$ denotes the number of off-diagonal operators
in the sequence. Note that since the expectation value in
Eq.~(\ref{part2}) is always equal to zero or one, the sign of a term
is negative only if $n_2$ is odd. This sign problem occurs (only) when
frustration is present and is the main topic of this paper. However,
for the discussion of the sampling procedures, in this section we
assume a positive definite expansion. We introduce the notation 
$|\alpha (p)\rangle$ for a propagated state 
\begin{equation}
|\alpha (p)\rangle =\prod_{i=1}^p H_{a_i,b_i}|\alpha\rangle ,
\label{prop}
\end{equation}
where for an allowed configuration $|\alpha (0)\rangle = |\alpha
(L)\rangle = |\alpha\rangle$ and the weight function corresponding to 
(\ref{part2}) is given by
\begin{equation}
W(\alpha,S_L)= {(J\beta)^n(L-n)! \over 2^n L!}.
\label{weight}
\end{equation}

Having established the framework we will proceed to describe the
procedures for importance sampling of the terms $(\alpha,S_L)$
according to the weight (\ref{weight}). The initial state can be a 
sequence of the form $(0,0)_1,(0,0)_2,\ldots, (0,0)_L$ (subscripts on 
the index pairs will sometimes be used to denote the position in $S_L$) 
and a random state $|\alpha\rangle$. An ergodic procedure for sampling the 
terms is achieved using two types of basic updates; a simple substitution of
single diagonal operators and the operator-loop update which involves
simultaneous updates of a number (in principle varying between $1$ and
$n$) of diagonal and off-diagonal operators.

The diagonal update is carried out by
traversing the operator-index sequence $S_L$ from beginning ($p=1$) to
end $(p=L)$.  Operator substitutions of the form
$(0,0)_p\leftrightarrow (1,b)_p$ are attempted where possible, while
the stored state $|\alpha \rangle$ is updated every time an
off-diagonal operator is encountered so that the state $|\alpha (p)
\rangle$ is available when needed. With the weight function (\ref{weight})
detailed balance can be seen to be satisfied if the acceptance
probabilities are taken to be
\begin{eqnarray}
P\left[(0,0)_p\rightarrow (1,b)_p \right]&=& \frac{M\beta\langle\alpha_b(p)
|H_{1,b}|\alpha_b(p)\rangle}{L-n} ,\\
P\left[(1,b)_p\rightarrow (0,0)_p \right]&=& \frac{L-n+1}{M\beta
\langle\alpha_b(p) |H_{1,b}|\alpha_b(p)\rangle}.
\end{eqnarray}
Note that the diagonal update changes the expansion power $n$ by $\pm 1$.
Off-diagonal operators cannot be introduced one-by-one because of the
periodicity condition $|\alpha (L)\rangle = |\alpha (0)\rangle$. Local
updates involving two operators can be used for this purpose,\cite{sse2} 
but are more complicated and far less efficient than the operator-loop
update,\cite{sse3} which is discussed next.

\begin{figure}[htb]
\centering
\epsfxsize=5.0cm
\leavevmode
\epsffile{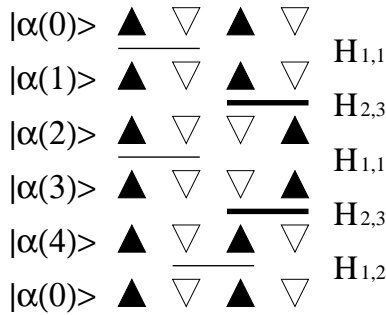}
\vskip4mm
\caption{Representation of a term in the SSE expansion of a four-site
antiferromagnet. Up and down spins are represented as solid and
open triangles, respectively. The horizontal bars indicate the
presence of diagonal (narrow lines) and off-diagonal (thick lines)
operators.}
\label{fig0}
\end{figure}

We use a largely pictorial description of the operator loops. First we 
consider the antiferromagnet. Note that in this case the only non-zero 
matrix elements of the bond operators are
\begin{eqnarray}
\langle \downarrow\uparrow|H_1|\downarrow\uparrow\rangle&=&
\langle \uparrow\downarrow|H_1|\uparrow\downarrow\rangle=1\nonumber ,\\
\langle \downarrow\uparrow|H_2|\uparrow\downarrow\rangle&=&
\langle \uparrow\downarrow|H_2|\downarrow\uparrow\rangle=1 ,
\label{elema}
\end{eqnarray}
i.e., they can act only on antiparallel spins.  An example of a term
in the expansion for a four-site antiferromagnet is depicted in
Fig.~(\ref{fig0}). This representation makes evident the close relationship
between the SSE expansion and the Euclidean path integral. An imaginary time 
separation $\tau$ corresponds to a distribution of propagations $\Delta p$ 
between states, centered around $\Delta p = (\tau/\beta)n$.\cite{sse1,irsse} 
We will for convenience here refer to the propagation as the time dimension.

The general idea \cite{evertz} behind the loop update is to flip a cluster 
of spins in the configuration in such a way that the weight is not changed.  
With the SSE method there will also have to be changes made to the 
operators acting on the spins, since otherwise operators $H_1$ or $H_2$ 
may act on parallel spins, resulting in zero-valued matrix elements. 
Since one of the states $|\alpha (p)\rangle$ and the operator sequence 
$S_L$ uniquely define the whole spin configuration, the SSE loops are in 
practice treated as loops of operators, the exact meaning of which will 
be made clear below. 

\begin{figure}[htb]
\centering
\epsfxsize=3.0cm
\leavevmode
\epsffile{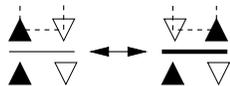}
\vskip4mm
\caption{Spin flip and accompanying operator exchange during a
loop update for an antiferromagnet. The dashed line indicates part
of the loop.}
\label{fig01}
\end{figure}

Consider one of the operators
$H_{1,1}$ in Fig.~(\ref{fig0}). It can be depicted as a ``vertex''
with four legs associated with spin states $\downarrow$ or
$\uparrow$. If we flip the upper left spin, a vanishing matrix element
results. But if we flip both upper (or lower) spins and also change
the operator type to off-diagonal $H_{2,1}$, an allowed matrix element
is generated; see Fig.~(\ref{fig01}). Using this idea we can form a
cluster of spins by choosing a random spin $S^z_i(p)$ in the
configuration and traversing up or down until we encounter an operator
(bond) acting on that spin, then switch to the second spin of the bond
and change the direction of traversing the list. Eventually we will
necessarily arrive back at the initial starting point, whereupon a
closed loop has formed. All the spins on this loop can be flipped if
the operators encountered are also switched [$(1,b) \leftrightarrow
(2,b)]$. Note that the same operator can be encountered twice, which
results in no net change of operator type (but the spins at all four
vertex legs are flipped).

\begin{figure}[htb]
\centering
\epsfxsize=7.0cm
\leavevmode
\epsffile{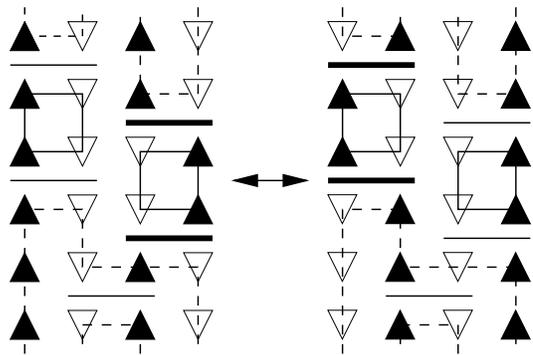}
\vskip4mm
\caption{The SSE space-time configuration of Fig.~1 uniquely divided up
into loops (left). The right hand configuration results from flipping the
loop indicated by the dashed line. Note the periodic boundary conditions
in the vertical (``time'') direction.}
\label{fig02}
\end{figure}

The whole configuration can be uniquely divided up into a set of loops, so 
that each spin belongs to one and only one cluster; see Fig.~(\ref{fig02}), 
where our example configuration has been divided up into three clusters. All 
loops can be flipped independently with probability $1/2$ --- 
in Fig.~(\ref{fig02}) we depict the result of flipping the largest loop 
of the example configuration. A full operator-loop update amounts to dividing
up the configuration into all of its loops which are flipped with probability 
$1/2$. The (random) decision of whether or not to flip a loop can be made 
before the loop is constructed, so that each loop has to be traversed only 
once. Spin ``lines'' $S^z_i(p)$, $p=0,\ldots ,n$, which are not acted upon 
by any of the operators in $S_L$ will not be included in any of the 
operator-loops. They correspond to  ``free'' spins which can be flipped 
with probability $1/2$. Such a line can also be considered a loop, and
then it will always be true that every spin $S^z_i(p)$ belongs to one
loop. Free spins appear frequently at high temperatures, when the total 
number of operators $n \alt N$, but are rare at low temperatures. 

Note that the spin states at the four legs of the operator-vertices
completely determine the full spin configuration, except for free spins
that happen not to be acted upon by any of the operators in $S_L$.
Hence, the operator-loop update can be carried out using only a linked list
of of operators, i.e., an array of vertices with four spin states and
associated pointers to the ``previous'' and ``next'' vertex associated with
the same spin.  The storage requirements and the number of operations 
needed for carrying out a full operator-loop update then scale as 
$\sim N\beta$ instead of $\sim N^2\beta$ if the full spin configuration 
were to be used. 

In a simulation we first make a full cycle of diagonal updates
in the sequence $S_L$ and then create the linked list of vertices in which 
the operator-loop updates are carried out. The vertex list is then mapped
back onto the sequence $S_L$ and the state $|\alpha (0)\rangle$. 
Alternatively, the linked list can be updated simultaneously with each 
diagonal operator substitution, so that it does not have to be re-created 
for each Monte Carlo step --- depending on the model studied there
may be significant differences in execution time between the two
approaches.

\begin{figure}[htb]
\centering
\epsfxsize=3.0cm
\leavevmode
\epsffile{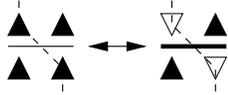}
\vskip4mm
\caption{Spin flip and accompanying operator exchange during a
loop update for a ferromagnet. The dashed line indicates part of
the loop.}
\label{fig04}
\end{figure}

\begin{figure}[htb]
\centering
\epsfxsize=7.0cm
\leavevmode
\epsffile{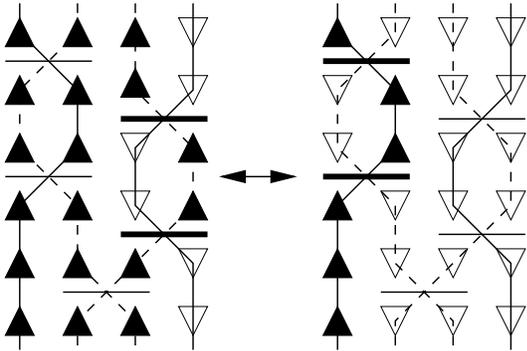}
\vskip4mm
\caption{SSE space-time configurations uniquely divided into loops
for the case of a ferromagnet. The left and right hand configurations
differ by flipping the loop indicated by a dashed line. }
\label{fig05}
\end{figure}

For the ferromagnet we can construct the loops in a similar
manner, but the non-zero matrix elements are now 
\begin{eqnarray}
\langle \downarrow\downarrow|H_1|\downarrow\downarrow\rangle&=&
\langle \uparrow\uparrow|H_1|\uparrow\uparrow\rangle=1, \nonumber\\
\langle \downarrow\uparrow|H_2|\uparrow\downarrow\rangle&=&
\langle \uparrow\downarrow|H_2|\downarrow\uparrow\rangle=1,
\label{elemf}
\end{eqnarray}
i.e., the off-diagonal operators act only on antiparallel spins, as before, 
whereas the diagonal ones can act only on parallel configurations. This
implies qualitative changes in the structure of the loops, as depicted in 
Fig.~(\ref{fig04}). If we again consider an operator $H_{1}$ and flip 
the upper left spin, we note that we need to flip the lower right 
spin and change the operator to $H_2$. Hence, instead of changing the direction
of traversing the configuration every time an operator is encountered we
now continue in the same direction after switching to the second spin of 
the bond. Any configuration can still be uniquely divided up into loops
that can be flipped with probability $1/2$. An example of a ferromagnetic
configuration with its loops is shown in Fig.~(\ref{fig05}).

Note that since the loops for the ferromagnet never change direction as
they go through the lattice, every single loop has to traverse each state 
$|\alpha (p)\rangle$ at least once. It follows that the number of sites $N$
is an upper bound of the number of loops. The antiferromagnetic loops,
on the other hand, traverse the lattice in both directions and the number 
of loops is therefore instead limited by the total number of operators 
$n \sim N\beta$. As a consequence of the change of direction, for the 
antiferromagnet the linked list of vertices must to be bi-directional, but 
for the ferromagnet it is sufficient to keep a singly-directional list.

The diagonal and operator-loop updates satisfy detailed balance and the
combination of them leads to ergodic sampling for a ferromagnet on any 
lattice, and for antiferromagnets on bipartite lattices --- for frustrated 
antiferromagnets there are complications, in addition to the sign problem, 
which will be discussed further in the next section. 
The operator-loop sampling is highly 
efficient, with integrated auto-correlation times typically less than one 
Monte Carlo step.

The loop construction described here relies on the rotational invariance
of the models, i.e., the fact that both the diagonal and off-diagonal matrix 
elements in Eqs.~(\ref{elema}) and (\ref{elemf}) are equal to one. 
For a general anisotropic
case the loops will lead to weight changes when flipped and must then be 
assigned ``a-posteriori'' acceptance probabilities which typically become 
small for large lattices at low temperatures. Other types of loops avoiding 
this problem have been proposed,\cite{sse3} but will not be discussed here.

\section{THE SIGN PROBLEM}

The notorious sign problem arises in stochastic sampling when the
function used to weight the different configurations in not positive
definite. A typical quantity that can be calculated by 
Monte Carlo (importance sampling) is of the form
\begin{equation}
\langle A\rangle =\frac{\sum_i A(x_i)W(x_i)}{\sum_i W(x_i)}=
\langle A(x)\rangle_W,
\end{equation}
where $W$ is the weight function and $A$ the measured quantity, which
both depend on the general coordinate $x$ of the configuration space
sampled. When the coordinates are sampled according to relative
weight, the desired quantity is simply the arithmetic average of the
measured function $A(x)$, as indicated by the notation $\langle
A(x)\rangle_W$ above. If the weight function is not positive definite,
the sampling can be done using the absolute value of the weight, and
the expectation value can be calculated according to 
\begin{equation}
\langle A\rangle =
\frac{\langle A(x)\mbox{s}(x)\rangle_{|W|}}
{ \langle s(x)\rangle_{|W|}},
\end{equation}
where $s(x)$ equals $\pm 1$, depending on whether the sign of the
weight function is positive or negative. For most models, where a sign
problem is present, the average sign $\langle s(x)\rangle_{|W|}$
decreases exponentially to zero as a function of inverse temperature
and system size, and the relative statistical errors of calculated
quantities increase exponentially.

A meron-cluster solution to the sign problem using loop updates of fermion
world-line configurations was recently proposed by Chandrasekharan
and Wiese.\cite{wiese} This approach is based on the idea that if it 
is possible to map every configuration with negative weight uniquely 
to a corresponding configuration with equal weight magnitude but opposite 
sign, then the partition function can be sampled without a sign problem
simply by not including any configuration which is a member of such
a canceling pair. In the meron-cluster algorithm, flipping a loop of
spins can lead to a sign change without change in the magnitude of
the weight, and such a ``meron'' hence identifies a canceling pair of 
configurations. Here we will present a similar approach within the SSE 
operator-loop method for frustrated quantum spins.

Let us consider the Heisenberg antiferromagnet discussed in the previous 
section. From Eq.~(\ref{part2}) we see that a configuration has negative 
weight if the total number $n_2$ of off-diagonal operators is odd. This 
can only be the case on frustrated lattices. As described in the previous 
section, any SSE configuration can be divided up uniquely into a number of 
loops. Flipping a loop interchanges the diagonal and off-diagonal operators, 
but leaves the total number of operators unchanged. It follows that the sign
will change if and only if a loop passing through an odd number of operators 
is flipped (two passes through the same operator is counted as two operators).
Since the total weight remains unchanged we have thus found
the desired mapping between positive and negative configurations (assuming
that there exist loops which change the sign when flipped, which in fact
is not always the case). In analogy with previous work we call such a 
sign-flipping loop a ``meron''. Let us now see how we can use this concept 
to measure observables.

As in Ref.~\onlinecite{wiese} we consider improved estimators that
average over all loop configurations. Denote the number of loops in
the system $N_L$. Since each loop can be in one of two states 
there is  a total of $2^{N_L}$ configurations that can be reached by 
flipping all combinations of the loops present. The improved estimate
therefore takes the form
\begin{equation}
\langle A\rangle =\frac{\langle\langle A(x)s(x)\rangle\rangle_{|W|}}
{\langle\langle s(x)\rangle\rangle_{|W|}},
\label{improved}
\end{equation}
where the double brackets denote an average over all the loop states
for each generated SSE configuration, e.g.,
\begin{equation}	
\langle\langle s(x)\rangle\rangle=
\left \langle\frac{1}{2^{N_L}}\sum_{l=1}^{2^{N_L}}s(x_l)\right\rangle.
\end{equation}
The general coordinate $x$ here refers to the SSE configuration space 
$(\alpha,S_L)$ and $x_l$ refers to one out of the $2^{N_L}$ possible
outcomes of ``flipping'' a number of loops.

Let us consider this average. Denote the state of a loop with
$\delta$, with two possible ``orientations'' $\delta\in
\{\uparrow,\downarrow\}$. Since flipping one loop does not affect any
other loops (in terms of their paths taken), 
the sign of a configuration factors according to
\begin{equation}
s(\delta_1,\delta_2,\ldots,\delta_{{N_L}})=
\prod_{i=1}^{N_L} s(\delta_i),
\end{equation}
where $s(\delta) =\pm s(\bar{\delta})$, where $\bar{\delta}$ denotes a
flipped loop, and the sign is negative for merons and positive
otherwise. Since flipping any meron leads to two terms that cancel, 
it follows that 
\begin{equation}
\frac{1}{2^{N_L}}\sum_{l=1}^{2^{N_L}}s(x_l) =\pm \delta_{n_M,0},
\end{equation}
where $n_M$ denotes the total number of merons. The sign in front
of the delta function is the ``inherent'' sign of the configuration,
independent of the loop orientation when there are no merons present.
This sign has to be positive for the meron solution to be applicable
in practice, and then the partition function can be sampled in the 
positive definite subspace of configurations with no merons.

Having found an expression for the denominator in Eq.~(\ref{improved})
we need to consider the numerator for cases of interest. SSE estimators for
a number of important operators have been discussed, e.g., in 
Ref.~\onlinecite{sse2}. The internal energy is given by
\begin{equation}
E=-\frac{1}{\beta}\langle n\rangle_W,
\end{equation}
where $n$ denotes the total number of operators in the sequence $S_L$.
This number is not affected by the loop updates, and hence it
follows that 
\begin{equation}
E=-\frac{1}{\beta}\frac{\langle\langle
n(x)s(x)\rangle\rangle_{|W|}} {\langle\langle
s(x)\rangle\rangle_{|W|}}=-\frac{1}{\beta}
\frac{\langle
n(x)\delta_{n_M,0}\rangle_{|W|}} {\langle\delta_{n_M,0}\rangle_{|W|}}.
\end{equation}
Assuming  that this sector has positive definite
weight we have therefore completely eliminated the sign
problem by restricting the simulation to the zero-meron
sector. The energy is then simply given by
\begin{equation}
E=-\frac{\langle n(x^0)\rangle_W}{\beta},
\end{equation}
where the superscript 0 indicates the restriction of the simulation to the 
zero-meron sector.

Next we will consider the magnetic susceptibility, 
\begin{equation}
\chi= \beta\left\langle \left(\sum_i S_i^z\right)^2\right\rangle/N = \beta \langle M ^2\rangle /N.
\end{equation} 
Since $M$ is conserved by the Hamiltonian its value is the same in all
propagated states; $M(p)=M(0)\equiv M$ $(p=1,\ldots ,n)$. In a configuration 
uniquely divided up into loops, every spin $S^z_i(p)$ belongs to one and only
one loop, if we count as loops also all ``lines'' of free spins, i.e., the 
spins $S^z_i(p)$, $p=1,\ldots, n$ for all sites $i$ which are not associated 
with any operator in the sequence (and therefore can be flipped). It follows
that the change in $M(p)$ when flipping a loop must be the same for
all $p$, and hence only loops that go through all states $|\alpha (p) 
\rangle$ (at least once) can change $M$ when flipped. We can therefore
introduce a loop magnetization $m_L$, which is simply equal to the sum 
of the spins traversed by the loop for an arbitrary $|\alpha (p) \rangle$.
In the estimator (\ref{improved}) corresponding to the susceptibility the 
numerator can hence be written as
\begin{equation}	
\left\langle\frac{1}{2^{N_L}}\sum_{l=1}^{2^{N_L}}(m_1(x_l)+ \ldots + 
m_{N_L}(x_l))^2 s(x_l)\right\rangle,
\label{loopsum}
\end{equation}
The magnetization on a loop always changes sign when a loop is flipped; 
the over-all sign $s(x_l)$ only changes sign when a meron is flipped. 
Therefore, in summing over all loops in (\ref{loopsum}), a non-zero value 
results only if the configuration has zero or two merons. The full 
susceptibility estimator therefore takes the form
\begin{equation}	
\chi = {\left\langle\sum_{l=1}^{2^{N_L}}|m_l|^2\delta_{n_M,0}+ 
2|m_{M_1}||m_{M_2}|\delta_{n_M,2}\right\rangle \over 
\langle\delta_{n_M,0}\rangle},
\label{susc1}
\end{equation}
where $M_1$ and $M_2$ are the indices of the loops corresponding to
merons in a two-meron configuration. Hence, unlike in the case of the
total energy, the sign problem has here not been completely
eliminated, since the zero- and two-meron configuration contribute $1$
and $0$, respectively, to the average sign. When the SSE configuration
volume grows the relative weight of the zero-meron sector should
diminish, leading to a decreasing average sign. Chandrasekharan and
Wiese stated that the statistical fluctuation in the improved
susceptibility estimator increases quadratically with $N\beta$, i.e.,
much slower than the conventional exponential increase.  They also
pointed out that this remaining sign problem can be solved by reweighting
the zero- and two-meron sectors with external weight factors $w(0)$
and $w(2)$. This changes the above formula to
\begin{equation}	
\chi = \frac{\left\langle\sum_{l=1}^{2^{N_L}}
|m_l|^2\delta_{n_M,0}w(2)+ 2|m_{M_1}||m_{M_2}|\delta_{n_m,2}w(0)\right
\rangle}{\langle\delta_{n_M,0}w(2)\rangle}
\label{eqrew}
\end{equation}
In the next section we will say more about reweighting.

As we have shown above, the meron concept within the SSE method
formally leads to exactly the same equations as in the world-line
simulations of fermion systems considered in
Ref.~\onlinecite{wiese}. The difference is only in the structure of
the meron itself; the fermionic meron changes the number of particle
permutations from even to odd, or vice versa, whereas the SSE meron in
the case of a frustrated spin system instead changes the number of
antiferromagnetic spin flips from even to odd or vice versa. Applying
the SSE operator-loop algorithm to a fermion system would lead to
merons of exactly the same kind as those existing within the
world-line framework, and, conversely, applying a world-line loop
algorithm to a frustrated spin system should lead to merons similar to
those discussed here (there are no diagonal operators in the
world-line configurations, but spin flip events correspond to the SSE
off-diagonal operators and can change from even to odd, or vice versa,
in loop updates). These similarities are not surprising, considering
the close relationship between the SSE expansion and the Euclidean
path integral.\cite{irsse}

\begin{figure}[htb]
\centering
\epsfxsize=6.0cm
\leavevmode
\epsffile{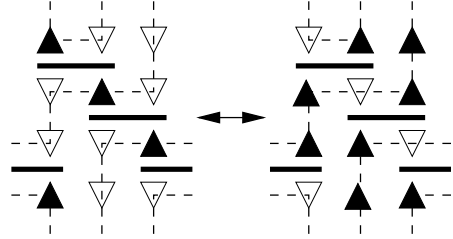}
\vskip4mm
\caption{Three off-diagonal operators acting on a triangle of spins.
The left and right hand configurations differ by flipping the loop
indicated by a dashed line. Note the periodic boundary conditions. }
\label{fig25}
\end{figure}

Now consider the application of the above ideas to the Heisenberg
model on a square lattice with nearest- and next-nearest-neighbor
couplings $J(1) > 0$ and $J(\sqrt{2}) > 0$ (antiferromagnetic). This
model has a sign problem since the total number $n_2$ of spin-flipping
operators in a configuration can be odd, e.g., three operators on a
triangle of spins, as shown in Fig.~(\ref{fig25}). Already this simple
example illustrates that the operator-loop algorithm discussed above
does not sample the full configuration space and that the meron
concept therefore cannot be applied to solving the sign problem. Since
all three bonds are antiferromagnetic, a loop will change direction
every time an operator is encountered. In order for the loop to close,
it therefore has to pass through an even number of operators and hence
flipping the loop cannot change the number $n_2$ of off-diagonal
operators from odd to even, or vice versa. This is illustrated in
Fig.~(\ref{fig25}), where the only effect of flipping the single loop
in the system is to flip all the spins; the operators remain
unchanged.  Hence, merons do not even exist within the operator-loop
algorithm for this model, and the sampling is non-ergodic. A local
update can in principle be used in combination with the operator-loops
in order to make the sampling ergodic. However, a configuration can
then have a negative sign (of which Fig.~(\ref{fig25}) is an example)
even though there are no merons present. The principal requirement of
positive-definiteness in the zero-meron subspace (which in this case
is the full space) is hence not fulfilled. A similar problem seems to
affect all models with frustration in all of the spin components. It
is possible that some other way of constructing the loops could remedy
this, e.g., by switching to some other, non-trivial basis in which the
SSE expansion could be carried out. Other ways proposed for
constructing loops in the standard basis considered here do not
uniquely define the set of loops \cite{sse3} and can therefore not
easily be used with the meron concept.

We have found one class of spin models for which the meron ideas can
be successfully applied to solve the sign problem: Heisenberg models
which are antiferromagnetic in the $xy$-plane but ferromagnetic along the
$z$-axis., i.e., the ``semi-frustrated'' model (\ref{subversivemodel}), 
which can be written as
\begin{equation}
H=- \sum_{b=1}^M \frac{J_b}{2} \left( H_{1,b}-  H_{2,b} \right)+ C,
\label{028}
\end{equation}
where the bond-indexed operators are given by
\begin{eqnarray}
H_{1,b}&=& 2\left(\hbox{$\frac{1}{4}$} + S_{i(b)}^zS_{j(b)}^z\right),\\
H_{2,b}&=& S_{i(b)}^+S_{j(b)}^- + S_{i(b)}^-S_{j(b)}^+.
\end{eqnarray}
On a non-frustrated lattice this model is equivalent to an isotropic
Heisenberg ferromagnet, since $n_2$ is always even and the sign in
front of the operators $H_{2,b}$ in Eq.~(\ref{028}) is
irrelevant as $(-1)^{n_2}=1$ in Eq.~(\ref{part2}) --- the sign can also
be transformed away by a spin rotation on one of the two sublattices.
On a frustrated lattice, on the other hand, $n_2$ can be odd (the
lattice is no longer bipartite so that the transformation mentioned
above does not remove all signs), and the system is no longer
equivalent to the isotropic ferromagnet. The model has a classical
two-fold degenerate ferromagnetic ground state, but at finite
temperatures the transverse spin components are frustrated, and the
behavior will be different from the isotropic ferromagnet.  When
simulated in the $z$-basis using standard algorithms the semifrustrated
model has a severe sign problem, but the zero-meron sector is positive-definite
and the SSE meron solution can be applied. The structure of the
operator-loops is the same as for the ferromagnet and the loop algorithm 
is therefore ergodic.

The meron solution can be implemented in several different ways and we 
briefly describe how it was done in this work: During the sequential 
diagonal updates the linked list, representing the loop structure, is 
updated simultaneously with each accepted diagonal update. The loops are 
numbered and information is stored on whether each loop is a meron or not. 
During an attempted diagonal update only the loops directly affected by 
the operator substitution are updated. This permits easy and fast checking 
of whether the number of merons in the system has changed or not. If the 
new number of merons is different form zero or two the update is rejected, 
whereas if the number of merons changes from $i$ to $j$ it is accepted with
probability $w(j)/w(i)$, where $i,j \in \{0,2\}$, and $w(i)$ is the 
re-weighting factor assigned to meron sector $i$.

\begin{figure}[htb]
\centering
\epsfxsize=3.0cm
\leavevmode
\epsffile{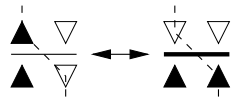}
\vskip4mm
\caption{Spin flip and accompanying operator exchange during a
loop update for model with a ferromagnetic $x$-component, but
antiferromagnetic $y$ and $z$-components. The dashed line indicates 
part of the loop.}
\label{fig06}
\end{figure}

Note that the sign problem for the semifrustrated model can also be very
easily transformed away by rotating the ferromagnetic component to the 
$y$-direction. The Hamiltonian then
takes the form
\begin{equation}
H=-\frac{J}{2} \sum_{b=1}^M \left( H'_{1,b}+  H'_{2,b} \right)+ C,
\end{equation}
where the bond-indexed operators are given by
\begin{eqnarray}
H'_{1,b}&=& 2\left(\hbox{$\frac{1}{4}$} - S_{i(b)}^zS_{j(b)}^z\right),\\
H'_{2,b}&=& S_{i(b)}^+S_{j(b)}^+ + S_{i(b)}^-S_{j(b)}^-,
\end{eqnarray}
and the fundamental spin-flips and operator exchange during a loop
update is shown in Fig.~(\ref{fig06}). Being able to work in both bases 
we can easily measure all components of the susceptibility. Our main
motivation for studying this model is to illustrate how the sign
problem can be removed in the $z$-basis. Nevertheless, we will also show 
some results calculated in the $x$-basis.

\section{REWEIGHTING}

An important technical aspect of the meron-solution is the reweighting
of the zero- and two-meron sectors, which was briefly mentioned in the
previous section. Eq.~(\ref{eqrew}) gives the correct estimator for
the susceptibility after reweighting, but it gives no information on
how to do the reweighting in practice.  How to determine the optimal
reweighting and whether reweighting changes the scaling of the
relative error are important questions to be considered in this
section.

\begin{figure}[htb]
\centering
\epsfxsize=7.25cm
\leavevmode
\epsffile{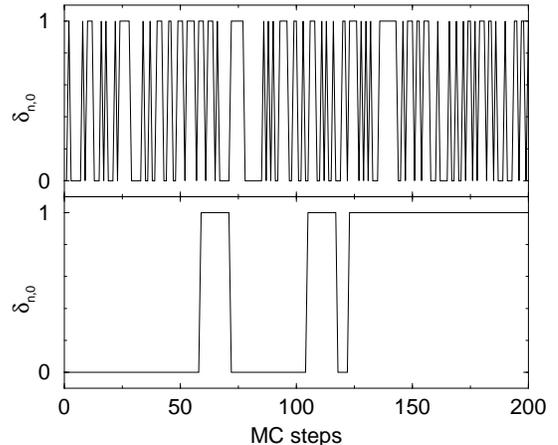}
\vskip1mm
\caption{Fluctuations of $n$ during a random process with two outcomes
($n=0,2$). The upper graph shows results for $W_0=0.5$ and $W_2=0.5$, 
while in the lower graph $W_0=0.01$ and $W_2=0.99$ and a reweighting factor
$W=99$ as used (resulting in $W_0'=W_2'=0.5$).}
\label{figa2}
\end{figure}

As a first, simple example of how reweighting affects the statistics of
a simulation we will discuss a simple random process. Consider a random 
variable $n$, which can take two different values, $0$ or $2$. Let $W_0$ 
and $W_2$ designate the probability for these two outcomes. The expected
fractions and standard deviations of these outcomes from $N$ random
selections are given by
\begin{eqnarray}
\langle \delta_{n,0}\rangle&=&W_0\pm\frac{\sqrt{W_0W_2}}{\sqrt{N}}\\
\langle \delta_{n,2}\rangle&=&W_2\pm\frac{\sqrt{W_0W_2}}{\sqrt{N}}.
\end{eqnarray}
We consider an expectation value of a form similar to the QMC
susceptibility, Eq.~(\ref{susc1});
\begin{equation}
\langle f \rangle = \frac{\langle \delta_{n,0}+\delta_{n,2}\rangle}
{\langle \delta_{n,0}\rangle}
=\frac{1}{W_0}\pm\frac{1}{W_0}\sqrt{\frac{W_2}{W_0}}\frac{1}{\sqrt{N}},
\label{eqrewf}
\end{equation}
with a relative standard deviation
\begin{equation}
\frac{\sigma_f}{f}=\sqrt{\frac{W_2}{W_0}}\frac{1}{\sqrt{N}}.
\end{equation}
This formula becomes valid for large $N$, when the standard deviation
is small. As $W_0$ decreases the standard deviation increases, but one
can reweight the two outcomes by assigning an additional weight $W$ to
the $n=0$ outcome such that a transition from $n=0$ to $n=2$ is
accepted with probability $1/W$, while a transition from $n=2$ to
$n=0$ is always accepted.  After such a reweighting the probabilities
of obtaining $n=0$ and $n=2$ are given by
\begin{eqnarray}
W_0'&=&\frac{W_0W}{W_0W+W_2}\\ W_2'&=&\frac{W_2}{W_0W+W_2},
\label{weights}
\end{eqnarray}
and $f$ is given by
\begin{equation}
\langle f \rangle = \frac{\langle \delta_{n,0}+W\delta_{n,2}\rangle}
{\langle \delta_{n,0}\rangle}.
\end{equation}
When calculating the standard deviation for this case we have to be
careful since the reweighting introduces correlations into the
system. This is clearly visualized in Fig.~\ref{figa2}, where in the
upper graph a series of independent outcomes with equal probability
($W_0=W_2=0.5$) are shown , while in the lower graph a case with
$W_0=0.01$ is shown with a reweighting factor of $W=99$ (leading to
$W_0'=W_2'=0.5$). 

Let us now calculate the standard deviations for this case. We are interested
in minimizing the standard deviation of a variable evaluated in a simulation
and also need to calculate its statistical error. In a standard MC simulation
one usually wants to calculate the average and the standard deviation of the 
average for some quantity $x$. This is typically achieved by dividing the 
run into a number of bins, $N$, and saving the average of $x$ for each bin. If 
the bins are statistically independent the final average and standard 
deviation can be calculated according to
\begin{equation}
\bar{x}=\frac{1}{N}\sum_{i=1}^{N}x_i
\end{equation}
and
\begin{equation}
\sigma_{\bar{x}}=\sqrt{\frac{\bar{x^2}-\bar{x}^2}{N}}.
\end{equation}
When studying the behavior of the standard deviation itself,  we also want 
to obtain an estimate of its accuracy.  This can be done by dividing the $N$
bins into $M$ sets containing $N/M$ bins each. For each set a standard
deviation $\sigma_{x}$ can be calculated according to
\begin{equation}
\sigma_{x}=\sqrt{\bar{x^2}-\bar{x}^2},
\end{equation}
where the bar denotes an average of the $N/M$ bins within the set.
The final standard deviation and its statistical fluctuation are then
given by
\begin{equation}
\bar{\sigma_{x}}= \frac{1}{M}\sum_{i=1}^{M}{\sigma_{x}}_i
\label{dev}
\end{equation}
and
\begin{equation}
\sigma_{\bar{\sigma_{x}}}=
\sqrt{\frac{\bar{\sigma_{x}^2}-\bar{\sigma_{x}}^2}{M}}.
\end{equation}
Eq.~(\ref{dev}) represents the standard deviation of the distribution
of the binned values $x$, and not the standard deviation of an average
of these. It does not decrease as the number of measurements $N$
is increased, but it is  dependent on the number of MC steps in each bin,
$N_{\mbox{\scriptsize bin}}$,  and will decrease as
$1/\sqrt{N_{\mbox{\scriptsize bin}}}$.  Hence it is important to state
the number of MC steps in the bins for which the deviation is
calculated.  The statistical error of this standard deviation,
$\sigma_{\bar{\sigma_{x}}}$, will, on the other hand, decrease as
$1/\sqrt{N}$.

\begin{figure}[htb]
\centering
\epsfxsize=7.25cm
\leavevmode
\epsffile{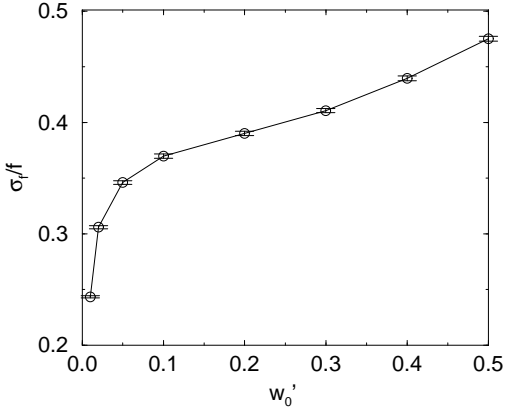}
\vskip1mm
\caption{
The relative standard deviation deviation of $f$ as a
function of $W_0'$, Calculations are performed over bins containing
$N_{\mbox{\protect\scriptsize bin}}=2000$ MC steps.
}
\label{figa3}
\end{figure}

\begin{figure}[htb]
\centering \epsfxsize=7.25cm \leavevmode \epsffile{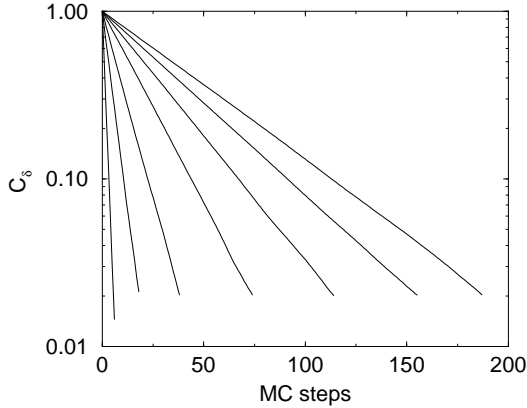}
\vskip1mm
\caption{Autocorrelation function, Eq.(\protect\ref{cdelta}), for
different values of $W_0'$. Shown are results for $W_0'=0.02$, $0.05$,
$0.1$, $0.2$, $0.3$, $0.4$, and $0.5$ (curves left to right).}
\label{figa4}
\end{figure}

In this manner we can calculate the standard deviation for $f$. In
order to show the standard deviation as a function of the reweighted
probability $W_0'$, Eq.~(\ref{weights}), can be inverted to express the
necessary weight factor that causes the average to change from $W_0$
to $W_0'$;
\begin{equation}
W=\frac{W_0'(1-W_0)}{W_0(1-W_0')}.
\label{weightsnew}
\end{equation}
Simulation results for the standard deviation of $f$ as a function of 
$W_0'$ is shown in Fig.~\ref{figa3}. We see that the reweighting actually 
increases the standard deviation. This is due to the rapidly increasing auto
correlation times. The autocorrelation function
\begin{equation}
C_{\delta}(t)=\frac{\langle\delta_{n,0}(i+t)\delta_{n,0}(i)\rangle
-\langle\delta_{n,0}(i)\rangle^2}{\langle\delta_{n,0}(i)^2\rangle
-\langle\delta_{n,0}(i)\rangle^2}
\label{cdelta}
\end{equation}
is shown in Fig.~\ref{figa4}, and one can see that the autocorrelation
times (inversely proportional to the slopes in Fig.~\ref{figa4}) are
proportional to $W_0'$. Notice that the longest autocorrelation times
are significantly shorter than the individual bins (consisting of 2000
MC steps) used above, a criteria for the analysis to be valid.

\begin{figure}[htb]
\centering
\epsfxsize=7.25cm
\leavevmode
\epsffile{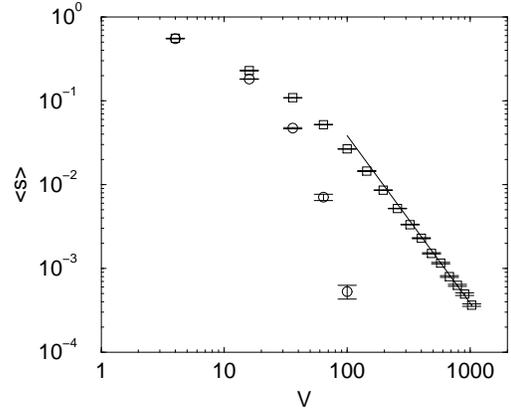}
\vskip1mm
\caption{The average sign as function of system volume at
temperature $T/J=1$. Shown are results of an unrestricted
simulation (circles), and a simulation restricted to the 0- and 2-meron
sectors without reweighting (squares). The line shows a slope
of -2.}
\label{fig2}
\end{figure}

This simple example seems to indicate that reweighting does not
decrease the statistical errors. However, in a standard Monte Carlo
simulation the measured quantities are not independent even with no
reweighting, and formula Eq.~(\ref{eqrew}) contains measured
quantities different from Eq.~(\ref{eqrewf}) considered above.
Therefore the reweighting will affect autocorrelation times
differently than in the above example, and reweighting can
actually decrease the standard error.\cite{wiese}

Using the above technique we can study how the relative error in the
susceptibility of the semi-frustrated model is affected by
reweighting.  An initial run without reweighting has to be done first
to determine the average sign $\langle
\delta_{n,0}\rangle=W_0$. Thereafter Eq.~(\ref{weightsnew}) can be
used to determine the desired weight factors. In Fig. \ref{fig2} the
average sign in a simulation of the semifrustrated model with
$J(1)=J(\sqrt{2})=J$ is shown as a function of lattice volume
$V=L\times L$ at a temperature $T/J=1.0$. For comparison we first
performed a standard simulation by sampling all meron sectors, which
leads to a severe sign problem with a sign that decreases
exponentially in system volume.  Next we sampled only the zero- and
two-meron sectors without reweighting, which dramatically increases
the average sign.  The scaling changes from exponential to quadratic
in the volume, as can clearly be seen from the graph.

\begin{figure}[htb]
\centering
\epsfxsize=7.25cm
\leavevmode
\epsffile{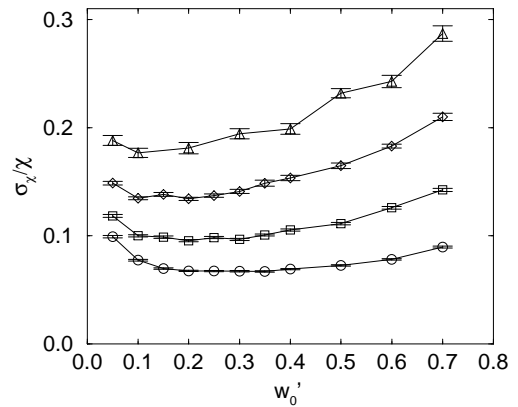}
\vskip1mm
\caption{Relative standard deviation of the $z$-susceptibility of the
semifrustrated model as a function of reweighting. Shown are results
for systems of linear size $8$ (circles), $12$ (squares), $16$ (diamonds),
and $20$ (triangles). The standard deviation is calculated for bins containing
$N_{\mbox{\protect\scriptsize bin}} = 10^3$ MC steps at temperature $T/J=1.0$.
}
\label{fig3}
\end{figure}

Having determined the average sign without reweighting we now use
Eq.~(\ref{weightsnew}) to determine the desired weight factors.
In Fig.~\ref{fig3} the standard deviation (\ref{dev}) of the 
susceptibility, calculated using bins containing $1000$ MC steps.
Results are shown for systems of linear size $N=8,12,16$ and 20 at 
temperature $T/J=1.0$.  Reweighting clearly helps to reduce the standard 
deviation, and there is a definite minimum in all these curves indicating 
an optimal reweighting. The optimally reweighted sign always appears to 
be less than 0.5, and decreases with decreasing sign (and increasing 
volume).

Having determined that there is an optimal reweighting we will next
consider whether reweighting changes the scaling with system size of the 
relative statistical error. Let us first consider how the standard deviation
scales with no reweighting.  Since the sign decreases quadratically in the
volume $V$ we can derive the scaling of the relative error in the
sign, under the ideal (and typically false) assumption that individual
measurements are completely independent.  Using that $s=s^2$ we arrive at
\begin{equation}
\frac{\sigma_s}{\bar{s}} = \frac{\sqrt{\bar{s^2}-\bar{s}^2}}
{\bar{s}\sqrt{N}} \sim \frac{1}{\sqrt{\bar{s}N}} =
\frac{V}{\sqrt{N}}
\label{linear}
\end{equation}
and this indicates that in this case the statistics needed actually
increases linearly in system volume, and not quadratically as stated
in Ref.~\onlinecite{wiese}.

\begin{figure}[htb]
\centering
\epsfxsize=7.25cm
\leavevmode
\epsffile{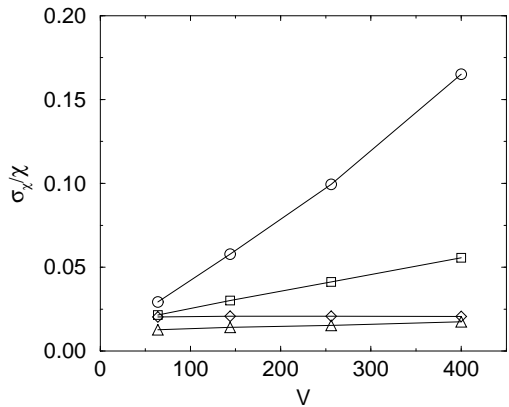}
\vskip1mm
\caption{ The standard deviation of the susceptibility as a function
of system volume V. Shown are: the $z$-component for the semifrustrated
model without (circles) and with (squares) optimal reweighting, 
the $x$-component for the semifrustrated model (diamonds) and the
rotationally invariant susceptibility of the ferromagnetic model (triangles).
Statistical errors are smaller than symbol size.}
\label{fig4}
\end{figure}

In order to study the scaling, the standard deviation for bins
containing $N_{\mbox{\scriptsize bin}}=10^4$ MC measurements of the 
susceptibility is shown in Fig.~\ref{fig4}.  Four susceptibilites are 
shown: the $z$-component for
the semifrustrated model without and with optimal reweighting, the
$x$-component for the semifrustrated model and the rotationally
invariant susceptibility of the isotropic ferromagnetic model. The two 
latter quantities can be obtained in simulations without sign problems, 
as discussed above.

Let us first consider the $z$-component of the susceptibility for the
semifrustrated model without and with optimal reweighting.  For {\it
both} cases the graph suggests a linear increase in relative error.
We have to keep in mind that Eq.~(\ref{linear}) does not have to be
valid, since there are autocorrelations in the simulation, and the
results in Fig.~\ref{fig4} do not exclude that the scaling changes
when approaching the thermodynamic limit (due to increasing
autocorrelation times), but both results do support an approximately
linear increase.  It appears that the reweighting in this case changes
only the prefactor of the volume scaling, but not the
exponent. Whether this means that reweighting has completely
eliminated the remaining sign problem \cite{wiese} is not clear (the
error remains larger than that for the ferromagnet susceptibility),
but it is clear that the reweighting reduces the standard deviations
by a significant factor. In any case, an algorithm that changes the
functional dependence of the size scaling of the statistics from
exponential to polynomial can be considered a solution to the problem.

The $x$-susceptibility of the semifrustrated model, which is evaluated
with an algorithm without sign problems, shows a constant standard deviation, 
which may be related to the fact that the susceptibility itself has converged
to its thermodynamic limit for these system sizes (see next section). For
the isotropic ferromagnetic model, the susceptibility still shows a
linearly increasing error, but as already noted the slope is much smaller 
than for the semifrustrated case.

This concludes our discussion of the reweighting technique. In future
work it would be interesting to explore how the optimal reweighting
can be determined directly from quantities measured during one single
test run, rather than by explicitly measuring the standard deviations
as we have done here.

\section{RESULTS}

In this section we will present results for the semifrustrated and
isotropic ferromagnetic models. 
We will demonstrate that it is feasible to obtain
accurate results for large systems in the $z$-basis by using the 
meron-solution. The main motivation for this study is to analyze the
meron-solution, and we will only briefly comment on the physics of
the semifrustrated model and how it differs from the isotropic 
ferromagnet. We will primarily consider the semifrustrated model
with $J(1)=J(\sqrt{2})=J$. To verify the correctness of our codes we have 
compared simulation results with exact diagonalization data for systems 
with $4\times 4$ spins.

\begin{figure}[!htb]
\centering
\epsfxsize=7.25cm
\leavevmode
\epsffile{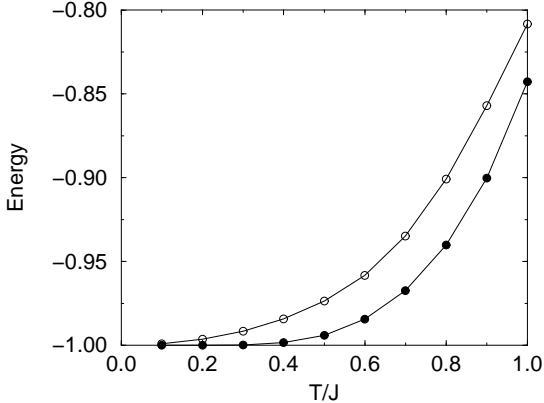}
\vskip1mm
\caption{Energy for the semifrustrated (filled circles) and
ferromagnetic (empty circles) model. Statistical errorbars are not
visible in the plot, and the curves have converged to the
thermodynamic limit.}
\label{fig5}
\end{figure}

In Fig.~\ref{fig5} the energy is shown for the semifrustrated and
ferromagnetic model. The statistical errorbars are smaller than the
symbol size and the results have converged to the thermodynamic limit.
The largest system size used had $128\times 128$ spins. It can be seen 
that thermal fluctuations and finite-temperature quantum fluctuations
more effectively destroy the ferromagnetic correlations   for the 
isotropic ferromagnet than for the semifrustrated model.

\begin{figure}[!htb]
\centering
\epsfxsize=7.25cm
\leavevmode
\epsffile{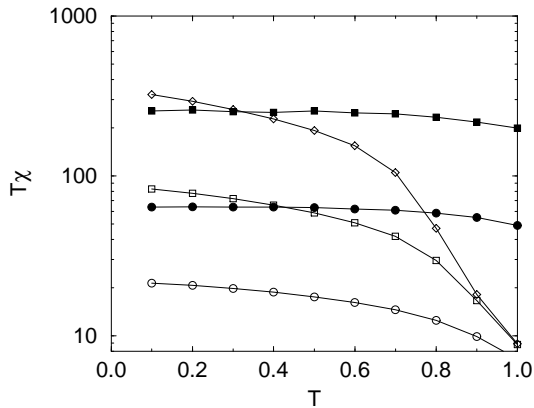}
\vskip1mm
\caption{The $z$-component of the susceptibility for semifrustrated
(filled symbols) and ferromagnetic model (empty symbols). The size
effects are shown for linear system sizes $N=16$ (circles), 32 
(squares) and 64 (diamonds, shown only for the ferromagnetic
model). Statistical errorbars are not visible on this scale.}
\label{fig6}
\end{figure}

The $z$-component of susceptibility is shown for both models in
Fig.~\ref{fig6}. The low-temperature susceptibility for finite-size
systems will approach $\beta N/4$ for the semifrustrated model, and
$\beta N/12$ for the ferromagnetic model (due to rotational averaging
in the latter case). This can be clearly seen from Fig.~\ref{fig6}, where 
the susceptibility is multiplied by temperature, so that a straight line 
at low temperatures indicates a Curie divergence. In the thermodynamic 
limit the uniform susceptibility should diverge exponentially for these 
models, but this can not be seen in Fig.~\ref{fig6} due to the strong 
finite-size effects.

\begin{figure}[!htb]
\centering
\epsfxsize=7.25cm
\leavevmode
\epsffile{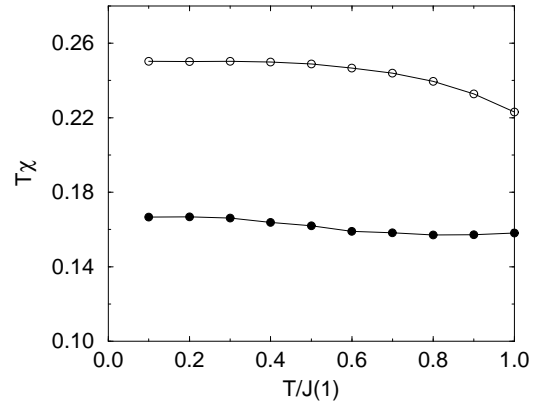}
\vskip1mm
\caption{The $x$-component of the susceptibility for the
semifrustrated model, at ratios  $J(\protect\sqrt{2})/J(1)=1$ (empty circles)
and $J(\protect\sqrt{2})/J(1)=0$ (filled circles).  Statistical errorbars are
not visible.}
\label{fig7}
\end{figure}

In Fig.~\ref{fig7} the $x$-component of the susceptibility for the
semifrustrated model is shown.  This quantity has converged to its
thermodynamic limit, and it exhibits a Curie divergence at low temperatures.
The ground state value is dependent on the next-nearest neighbor coupling 
$J(\sqrt{2})$, as is clearly demonstrated by plotting two different ratios 
$J\sqrt{2}/J(1)=1$, and $J\sqrt{2}/J(1)=0$ (only nearest neighbor 
interactions) in Fig.~\ref{fig7}.

\section{CONCLUSION}

We have studied a recently introduced meron-solution\cite{wiese} to
the sign problem within the SSE method. We investigated the sign
problem arising in frustrated spin systems and showed that the
meron-solution can be a applied to a particular model. The problems
arising when applying the meron solution to general models of frustrated 
spins were discussed. We found that loop algorithm typically are not 
ergodic and merons do not exist. The sign problem then persists.

For models where the meron solution works we showed that the sign
problem can be completely eliminated for certain variables and largely
eliminated for other variables. This total and partial elimination
comes form a mapping of positive- to negative-weight contributions and
involves no approximation. For the variables where the sign problem is
almost eliminated the statistical errors can be reduced using a
reweighting technique.\cite{wiese} We study how the relative statistical
error behaves as a function of lattice size with and without reweighting,
and showed that the reweighting does not change the scaling behavior
but typically significantly reduces the over-all magnitude of the 
fluctuations. 

It is evident that the meron solution 
suffers from the same problem in both frustrated spin and fermionic 
systems --- it is confined to a few special cases. It would be of 
great importance to be able to extend it to more general cases, e.g., by 
working in basis where the required loop structure appears. The possibility 
of finding such bases should be explored. 

\section{ACKNOWLEDGMENTS}

The research was supported by NSF Grants No. DMR-9629987, and
DMR-9712765. P.H. acknowledges support by Finska Vetenskaps-Societeten
and Suomalainen Tiedeakatemia.

\end{document}